\documentclass[a4paper]{jpconf}

\usepackage{hyperref}
\usepackage{amsfonts}
\usepackage{graphicx}
\usepackage{color,bm}

\begin{document}

\title{Where do moving punctures go?}
\author{Mark Hannam, Sascha Husa, Bernd Br\"{u}gmann, Jos\'e A. Gonz\'alez, Ulrich Sperhake}
\address{Theoretical Physics Institute, University of Jena, 07743 Jena, Germany}
\author{Niall \'O~Murchadha} 
\address{Physics Department, University College Cork, Ireland}

\date{\today}

\begin{abstract}
Currently the most popular method to evolve black-hole binaries is the ``moving puncture''
method. It has recently been shown that when puncture initial data for a Schwarzschild 
black hole are evolved using this method, the numerical slices quickly lose contact with the 
second asymptotically flat end, and end instead on a cylinder of finite Schwarzschild
coordinate radius. These slices are stationary, meaning that their geometry does not evolve
further. We will describe these results in the context of maximal slices, and 
present time-independent puncture-like data for the Schwarzschild spacetime.
\end{abstract}

\section{Introduction}

Recent breakthroughs in numerical relativity have made black-hole binary
evolutions routine for many research groups. The easiest
method to implement, and currently the most popular, is the ``moving puncture''
approach. Here one begins with initial data that possess a Brill-Lindquist
wormhole topology \cite{Brill63} and each asymptotic end is
compactified to a single point (puncture) on $R^3$ at the price of a
coordinate singularity. 
Punctures are technically appealing because they represent black holes
on $R^3$ without excision, the initial-data slices avoid the curvature singularity
of each black hole, and it is well understood how to construct 
puncture initial data for any number of boosted, spinning black holes 
\cite{Brandt97b,Dain02b}. 

The numerical evolution of puncture data is not so well understood. Successful
numerical techniques have been found only by trial and error. These include the
choice of formulation of the evolution equations, gauge conditions and, most
importantly, the treatment of the coordinate singularity at each puncture. Early
``fixed puncture'' evolutions \cite{Bruegmann97,Alcubierre00b} factored 
the singularity into a fixed analytically prescribed conformal
factor and a regular function that is evolved numerically, and used gauge
conditions that prevented the punctures from moving 
across the grid. These methods met with some success, but long-term stable
evolutions of general configurations of (especially orbiting) black-hole binaries
were not achieved. Puncture evolutions received very little analytic attention, but
it is interesting for the results we will describe here that the only detailed analytic 
study of the properties of fixed-puncture evolutions \cite{Reimann:2003zd,Reimann:2004pn1}, 
applied to the Schwarzschild spacetime, showed that such evolutions would never reach a 
stationary state: nontrivial evolution of the slices of Schwarzschild would continue
indefinitely.

Recently two groups~\cite{Campanelli:2005dd,Baker:2005vv} independently
introduced similar methods to deal with the puncture singularities. The singular 
conformal factor $\psi$ is evolved without any analytic assumptions 
(by evolving a new variable, either $\chi = \psi^{-4}$ 
\cite{Campanelli:2005dd} or $\phi = \ln \psi$ \cite{Baker:2005vv}), and the punctures
are able to move on the numerical grid. These methods have met with spectacular 
success, and the ``moving puncture'' method is now the most popular method
for evolving black-hole binaries --- of the eight codes reported to have 
successfully performed long-term evolutions of several orbits \cite{Pretorius:2005gq,
Campanelli:2005dd,Baker:2005vv,Herrmann:2006ks,Sperhake:2006cy,
Scheel:2006gg,Bruegmann:2006at,Pollney:NFNR}, six use moving 
punctures.

A number of questions were raised after the first moving-puncture results were
published. What happens to the punctures during the evolution?  Do 
they continue to represent compactified infinities? Does the evolution
reach a final, stationary state, or do gauge dynamics persist, as in the 
fixed-puncture case? And, crucially, does the method accurately describe 
the spacetime, or does it rely on numerical errors near the under-resolved
punctures, implying that it may fail when probed at higher resolutions or 
for longer evolutions?

These questions were answered by Hannam {\it et al} in \cite{Hannam:2006vv}, 
who studied moving-puncture evolutions of the Schwarzschild spacetime. 
The answers were, in short, that the evolution quickly reaches a stationary 
state. The slices no longer extend to the other asymptotically flat end, 
but instead end on a surface of finite Schwarzschild coordinate radius. 
For the version of 1+log slicing that the authors considered, the stationary 
slice ends at $R = 1.3M$. This point represents a throat and, although at a
finite Schwarzschild coordinate radius, it is an infinite proper distance from the
horizon. The authors also solve the stationary Einstein equations in spherical 
symmetry with the 1+log slicing condition, and show
that the numerical evolution accurately reproduces this solution, demonstrating
that the moving-puncture method is capable of accurately describing a 
black-hole spacetime, and the region near the ``puncture'' is not
under-resolved. 

In this article we will describe these effects in the context of
slices of Schwarzschild that are maximal in the initial and final states. 
The advantage of considering maximal slices
is that analytic closed-form expressions exist for these initial and final 
states. We may also easily construct time independent initial data
for a Schwarzschild black hole, which allow
us to study in more detail the numerical accuracy of the moving-puncture
method.

\section{Puncture evolutions of the Schwarzschild spacetime}

Following the standard 3+1 decomposition of Einstein's equations \cite{Arnowitt62,
York79} we write the spacetime metric as \begin{equation}
ds^2 = - \alpha^2 dt^2 + \gamma_{ij} \left( dx^i + \beta^i dt \right) \left(dx^j +
  \beta^j dt \right),
  \end{equation} where $\alpha$ is the lapse function, $\beta^i$ is the shift vector and 
$\gamma_{ij}$ is the spatial metric on one slice.
The data on each time slice are $(\gamma_{ij}, K_{ij})$, where the extrinsic curvature
is given by \begin{equation}
K_{ij} = \frac{1}{2\alpha} \left( \nabla_i \beta_j + \nabla_j \beta_i -
  \partial_t \gamma_{ij} \right). 
\end{equation} The data can be further decomposed by conformally relating the spatial
metric $\gamma_{ij}$ to a background metric $\tilde{\gamma}_{ij}$ via a conformal factor
$\psi$, and providing conformal weights to the trace and tracefree parts of the extrinsic 
curvature \cite{York79}, 
\begin{eqnarray*}
\gamma_{ij} & = & \psi^4 \tilde{\gamma}_{ij} \\
K_{ij}      & = & \psi^{-2} \tilde{A}_{ij} + \frac{1}{3} \gamma_{ij} K.
\end{eqnarray*} These data must satisfy constraint equations on each slice,
and the data on consecutive slices are related by evolution equations.  

It is in this context that puncture initial data are usually constructed. We
will restrict ourselves 
to the Schwarzschild solution. The standard Schwarzschild metric is \begin{equation}
ds^2 = - \left( 1 - \frac{2M}{R} \right) dt^2 + \left( 1 - \frac{2M}{R}
\right)^{-1} dR^2 + R^2 (d\theta^2  
+ \sin^2 \theta d \phi^2 ). 
\end{equation} If we make the coordinate transformation $R = \psi^2 r$, where
\begin{equation} 
\psi = 1 + \frac{M}{2r},
\end{equation} the Schwarzschild metric becomes \begin{equation}
 ds^2 = -\left( \frac{1 - \frac{M}{2r}}{1 + \frac{M}{2r}} \right)^2 dt^2 +
 \left( 1 + \frac{M}{2r} \right)^4 \left( dr^2 + r^2 d \Omega^2 \right).
 \end{equation} In these ``isotropic'' coordinates, the spatial metric takes the
 form $\gamma_{ij}  
 = \psi^4 \tilde{\gamma}_{ij}$, and the background spatial metric is the flat
 metric, $\tilde{\gamma}_{ij} = f_{ij}$. 
 
One important feature of isotropic coordinates is that they do not reach the
physical singularity at $R = 0$. For large $r$ we see that $R \rightarrow
\infty$, but for small $r$ we also see that $R \rightarrow \infty$. There is a
minimum of $R = 2M$ at $r = M/2$. We now have two copies of the space outside
the event horizon, $R > 2M$, and the two spaces are connected by a wormhole
(Einstein-Rosen bridge) with a throat at $R = 2M$. We refer to the second
asymptotically flat end (at 
$r = 0$) as the puncture. Schwarzschild $R$ as a function of isotropic $r$ is
shown in Figure \ref{fig:RvsX0}.

The Schwarzschild metric in isotropic coordinates is explicitly time
independent: if we evolve $\gamma_{ij}$ and $K_{ij}$ 
using the lapse and shift $\alpha = (1 - M/2r)/(1+M/2r)$, $\beta^i = 0$, we
will find that $\gamma_{ij}$ and $K_{ij}$ do 
not change. This is not to say that nothing will happen in a {\it numerical}
evolution: the lapse function is negative for $r < M/2$, and negative lapses
usually lead to numerical instabilities. However, the Schwarzschild metric in
isotropic coordinates is analytically time independent.  

We are free to slice the spacetime in some other way, i.e., to make a
different choice of lapse and shift. This will lead to a nontrivial evolution
of $(\gamma_{ij},K_{ij})$. The spacetime remains the stationary Schwarzschild
spacetime, but the coordinates will no longer be time independent. 

One simple choice for initial lapse and shift is $\alpha = 1$ and $\beta^i =
0$. If we maintain these choices
throughout the evolution, however, the slices will hit the singularity in finite
time, and we therefore choose to evolve the gauge in such a way that the
slices continue to avoid the singularity. Here we will choose the gauge
conditions popular in evolutions of black-hole binary puncture data. The lapse
is evolved with one of two variants of 1+log slicing \cite{Bona97a}, \begin{eqnarray} 
\partial_t \alpha & = & - 2 \alpha K \label{eqn:log} \\
\partial_t \alpha & = & - 2 \alpha K + \beta^i \partial_i \alpha, \label{eqn:logwithshift}
\end{eqnarray} and we evolve the shift vector with a $\tilde{\Gamma}$-freezing
condition \cite{Alcubierre02a}, \begin{eqnarray}
\partial_t \beta^i & = & \frac{3}{4} B^i,  \\
\partial_t B^i & = & \partial_t \tilde \Gamma^i - \eta B^i, \label{eqn:gammadriver} 
\end{eqnarray} where $\tilde{\Gamma}^i = - \partial_j \label{eqn:gammadriver2} 
\tilde{\gamma}^{ij}$. The lapse choice (\ref{eqn:logwithshift}) was analyzed
in \cite{Hannam:2006vv}, while here we will consider (\ref{eqn:log}). This
choice has the attractive property that a stationary state will be maximal
($2\alpha K = 0 \Rightarrow K = 0$). 

\subsection{Numerical slices}

Now let us consider evolving isotropic Schwarzschild data with the gauge
choices (\ref{eqn:log}), (\ref{eqn:gammadriver}) and
(\ref{eqn:gammadriver2}). Most current work 
uses the BSSN reformulation of the 3+1 equations \cite{Shibata95,Baumgarte99},
and the BSSN system has also been used to obtain the results 
shown here, using the BAM code \cite{Bruegmann:2006at}. (See also the articles
by Sperhake and Gonzalez in these proceedings.) However, the BSSN
system is irrelevant to the evolution of the geometry. Only gauge conditions
with the properties of (\ref{eqn:log}) - (\ref{eqn:gammadriver}) are
necessary. (That isn't to say that some 
other formulation will be numerically stable; but we are interested in how the
geometry evolves due to the gauge conditions, not the numerical properties of
a particular evolution system.) 

Having said that, the one element of the BSSN equations that must be singled
out if we are to discuss ``fixed'' versus ``moving'' punctures is the
treatment of the conformal factor, $\psi$, which is an independent evolution
variable in the BSSN system. The apparent problem with the conformal factor is
that it diverges at the puncture, and will presumably behave badly if evolved
directly. In ``fixed puncture'' evolutions one avoids this problem by writing
the initial isotropic Schwarzschild conformal factor as \begin{equation} 
\psi = \left(1 + \frac{M}{2r} \right) f,
\end{equation} where $f = 1$ at the beginning of the evolution, and then
evolving only $f$ (or, rather, $\ln f$), and leaving the divergent part
fixed. The method is tailored so that the wormhole topology remains throughout the
evolution. This means that even in black-hole binary simulations, where the
black holes have physical motion, the 
punctures are kept fixed on the grid \cite{Bruegmann97,Alcubierre00b}. 

It was pointed out in \cite{Reimann:2003zd} that Schwarzschild evolutions in
such a setup never reach a stationary state. One way of seeing this is
that if they did, the lapse would have to pass through zero at the throat, and
the slicing choices used in puncture evolutions preclude this. The same
problem was noted in attempts to construct quasi-equilibrium puncture data for
black-hole binaries with an everywhere positive lapse \cite{Hannam:2003tv}.

Now consider the ``moving-puncture'' method. Here one does not assume anything
about the form of $\psi$ during the evolution. We simply evolve either $\phi =
\ln \psi$ \cite{Baker:2005vv} or $\chi = \psi^{-4}$ \cite{Campanelli:2005dd}, both
of which behave sufficiently well at the puncture to allow stable
evolutions. (This may be a surprise for the choice $\phi = \ln \psi$, which
diverges logarithmically at the puncture, but it turns out that this
divergence is not strong enough to break the method.)

\begin{figure}
\centering
\includegraphics[angle=0,clip,width=.5\textwidth]{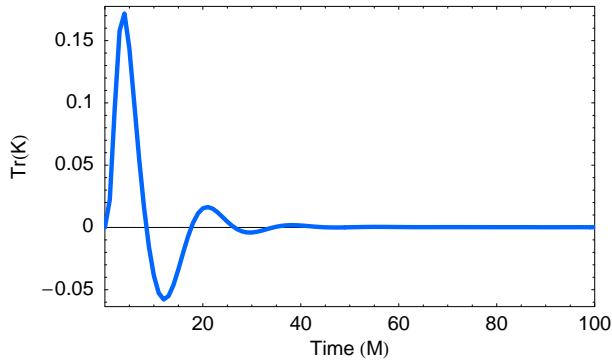}
\caption{The value of $K$ on the horizon $R = 2M$ as a function of time, for a
  moving-puncture evolution of Schwarzschild initial data. In this case the
  evolution reaches a stationary state.}
\label{fig:MaxK}
\end{figure}

Figure \ref{fig:MaxK} shows the value of $K$ at the horizion ($R = 2M$) during
a moving-puncture evolution. On the initial slice $K = 0$. As the evolution
progresses $K$ varies, but settles back to a 
maximal slice after about $40M$, and remains there. After this time a check of
other invariant quantities reveals that the evolution has reached a stationary
state.

\begin{figure}
\centering
\includegraphics[angle=0,clip,width=.5\textwidth]{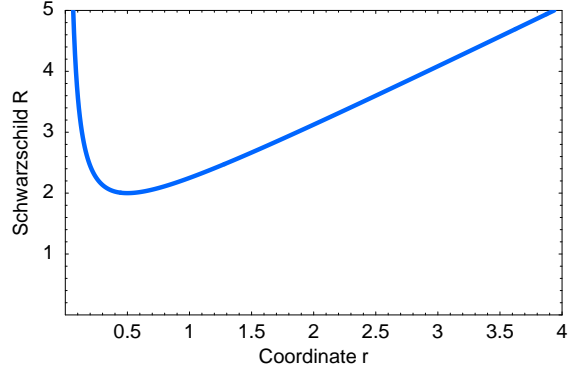}
\caption{The Schwarzschild radial coordinate $R$ as a function of numerical
  distance $r$ for the initial data.}
\label{fig:RvsX0}
\end{figure}

Now let us look at the evolution of the topology. Initially, the slice
connects two asymptotically flat ends. This is clearly seen in Figure \ref{fig:RvsX0},
which shows the Schwarzschild radial coordinate $R$ as a function of numerical
radial coordinate $r$. 

\begin{figure}
\centering
\includegraphics[angle=0,clip,width=.4\textwidth]{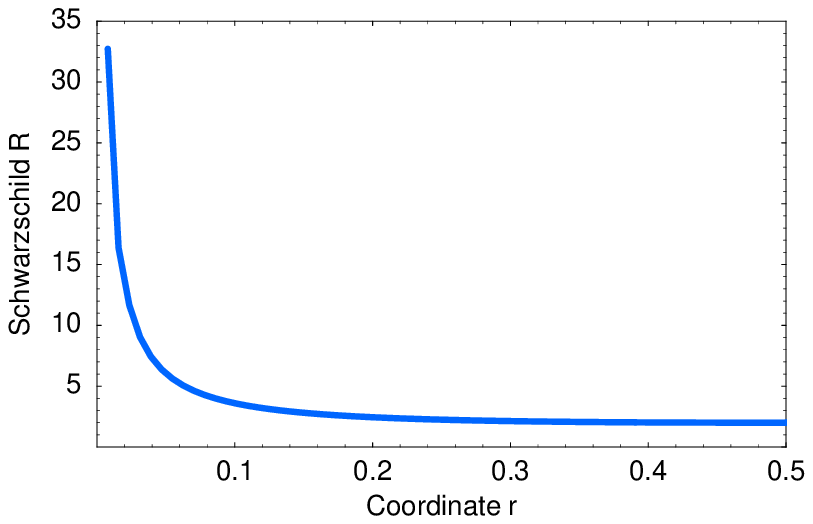}
\includegraphics[angle=0,clip,width=.4\textwidth]{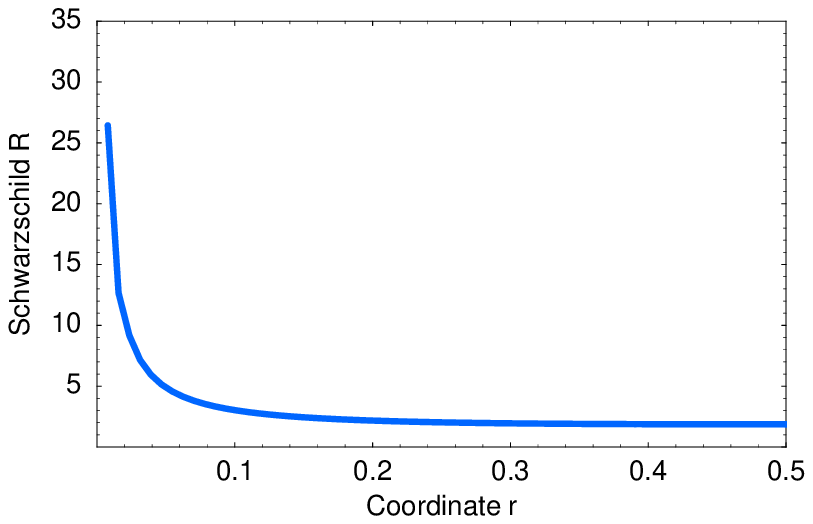}
\includegraphics[angle=0,clip,width=.4\textwidth]{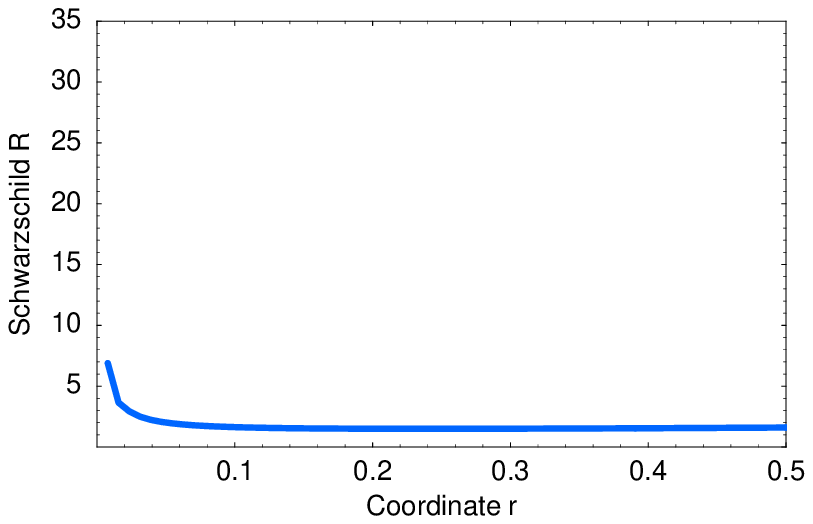}
\includegraphics[angle=0,clip,width=.4\textwidth]{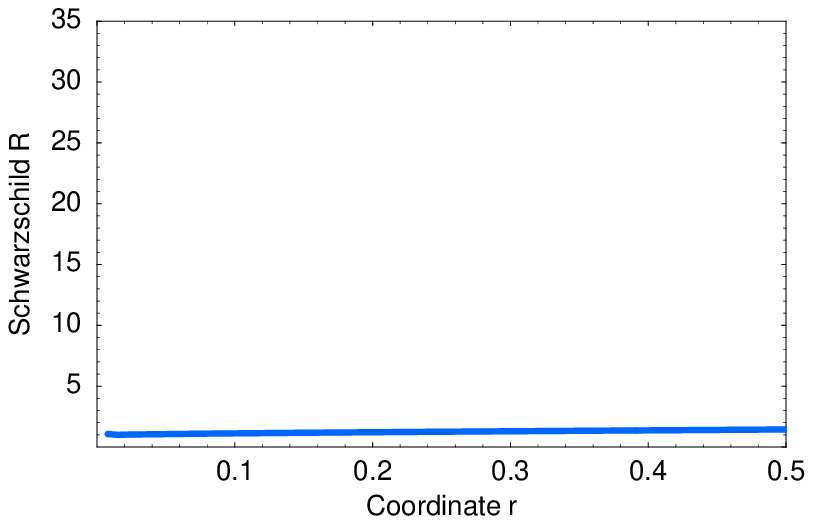}
\caption{The Schwarzschild radial coordinate $R$ as a function of numerical
  distance $r$ at times $T = 0,1M,2M,3M$ in the top-left, top-right,
  bottom-left and bottom-right figures. We see that the numerical slice loses
  contact with the second asymptotically flat end.}
\label{fig:RvsXT}
\end{figure}

\begin{figure}
\centering
\includegraphics[angle=0,clip,width=.5\textwidth]{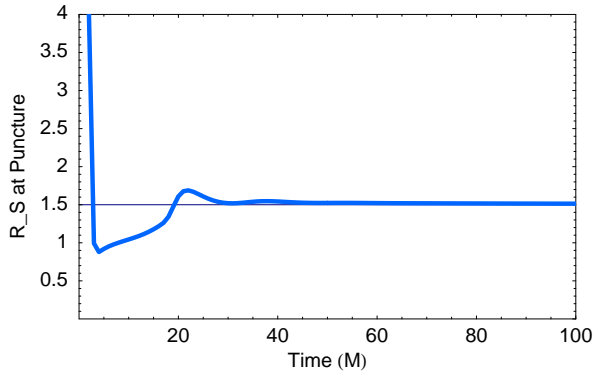}
\caption{The Schwarzschild radial coordinate $R$ at the puncture $r=0$ as a
  function of time. The end of the slice settles at $R = 3M/2$ after about
  40$M$ of evoluution.}
\label{fig:RvsXallT}
\end{figure}

Figure \ref{fig:RvsXT} shows $R(r)$ at times $t = 0,1,2,3M$ in a numerical evolution. We
see that the numerical slice quickly loses contact with the second
asymptotically flat end. The slice initially extends to $R \rightarrow \infty$
as $r \rightarrow 0$, but during evolution quickly retracts to a finite value
of the Schwarzschild radial coordinate. Figure \ref{fig:RvsXallT} shows the
value of $R$ at $r = 0$ as a function of time. After some complicated
oscillation, the end of the numerical slice settles at $R = 3M/2$, again after
about $40M$. 

In order for the slice to end at $R = 3M/2$, the nature of the singularity in
the conformal factor $\psi$ must change. In the initial data $\psi$ diverges as
$1/r$ at the puncture. By the time the evolution reaches a stationary state,
$\psi$ diverges as $1/\sqrt{r}$. In fact, near the puncture it will look like
$\psi \sim \sqrt{3M/2r}$ and $R = \psi^2 r \rightarrow 3M/2$. 
In the case of the modified 1+log
slicing (\ref{eqn:logwithshift}) studied in \cite{Hannam:2006vv}, the slice
ends at $R = 1.3M$, and $K \neq 0$. The numerical and analytic descriptions of
that slice were found to be in excellent agreement.

\subsection{Analytic maximal slices}

We may understand this result by comparison with the analytic result in the
1973 paper of Estabrook {\it et al} \cite{Estabrook73}. That paper contains an
analytic maximal slicing of the Schwarzschild spacetime for all time. The $t
\rightarrow 0$ limit of the Estabrook {\it et al} slicing is \begin{eqnarray*}
\gamma_{rr} & = & \left(1 - \frac{2M}{R}\right)^{-1}  \\
\beta^r & = & 0 \\
\alpha & = & 1.
\end{eqnarray*} If we transform to isotropic coordinates, we have precisely
the initial data that were used in our numerical evolution. Our final
stationary state is also a maximal slice, and therefore should represent (at
least part of) the $t \rightarrow \infty$ Estabrook {\it et al} solution,
which is \begin{eqnarray}
\gamma_{rr} & = & \left( 1 - \frac{2M}{R} + \frac{C^2}{R^4} \right)^{-1},
\label{eqn:EstaG} \\
\beta^r & = & \frac{\alpha C}{R^2}, \\
\alpha & = & \sqrt{1 - \frac{2M}{R} + \frac{C^2}{R^4}}, \label{eqn:EstaN}
\end{eqnarray}
with $C = 3\sqrt{3}/4$. In this solution we see that the lapse is zero at $R =
3M/2$: the slice 
ends there, just as in our numerical evolution! In a Kruskal diagram, the
throat is an infinitely long 3-cylinder \cite{Estabrook73}.
The same effect was seen in the numerical evolutions by Estabrook {\it et al}
in 1973, with a 1D code that required only 25 grid points. The remarkable
feature of the moving-puncture technique is that it finds this
time-independent slice (in slightly different coordinates) with the cylinder
at $R = 3M/2$ located at the origin of the numerical coordinate system, and
finds analogous slices in general configurations of multiple black holes. As
attractive as such an evolution system is, all of the tools of the
moving-puncture method were motivated by other considerations, and there was
no deliberate attempt to ``find'' slices like (\ref{eqn:EstaG})-- (\ref{eqn:EstaN}).
In the next section we will present Schwarzschild puncture initial data that
{\it are} motivated by the time-independent analytic solution.

\section{``Cylindrical'' initial data}

Motivated by the analytic Estabrook {\it et al} solution, we may attempt to
construct stationary initial data, i.e., data that do not change when
evolved with our 1+log and $\tilde{\Gamma}$-driver gauge conditions. 

To do this in a way that is most transparently related to the moving-puncture
method, we can conformally relate the $t \rightarrow \infty$ Estabrook {\it et
  al} spatial metric to a flat metric with a conformal factor that behaves as
$\psi \sim \sqrt{3M/2r}$ at the puncture. This conformal factor must also
behave as $\psi \sim 1 + M/2r$ as $r \rightarrow \infty$, in order to give the
correct ADM mass of the spacetime. We choose an ansatz that satisfies these
two conditions, and add to it a correction function such that the Hamiltonian
constraint is satisfied, in analogy to the standard puncture method for
black-hole binary initial data \cite{Brandt97b}. The ansatz we choose is
\begin{equation}
  \psi = \frac{2A(1 + r + 2r^2) + \sqrt{r}\left[2r^2(1+r) +
      M(1+r^2)\right]}{2\sqrt{r} (1 + r + r^2 + r^3)} + u, \label{eqn:NumPsi}
\end{equation} where $A = \sqrt{3/2}$ and $u$ is the correction function that
we solve for numerically. Both the flat-space laplacian of $\psi$ and the
function $\psi^{-7} \tilde{A}_{ij} \tilde{A}^{ij}$ in the source term of the Hamiltonian
constraint diverge as $r^{-3/2}$ at the puncture, but by multiplying the
entire Hamiltonian constraint by $r^{3/2}$ we are left with an equation in
which all terms are regular everywhere. The resulting elliptic equation for
the correction function $u$ can be solved in this case with a 1D code. 

After the first version of these proceedings were put online, Baumgarte and
Naculich \cite{Baumgarte:2007ht} pointed out that the Hamiltonian constraint
can be solved analytically for the $t \rightarrow \infty$ Estabrook metric,
at least in terms of the Schwarzschild coordinate $R$, $$
\psi^2 = \left( \frac{4 R}{2 R +  M + (4 R^2 + 4 M R + 3 M^2)^{1/2} }\right)
\left(\frac{8 R +  6 M + 3 ( 8 R^2 + 8 M R + 6 M^2 )^{1/2} }{(4 + 3
    \sqrt{2})(2 R - 3 M) }\right)^{1/\sqrt{2}}.
$$

\begin{figure}
\centering
\includegraphics[angle=0,clip,width=.45\textwidth]{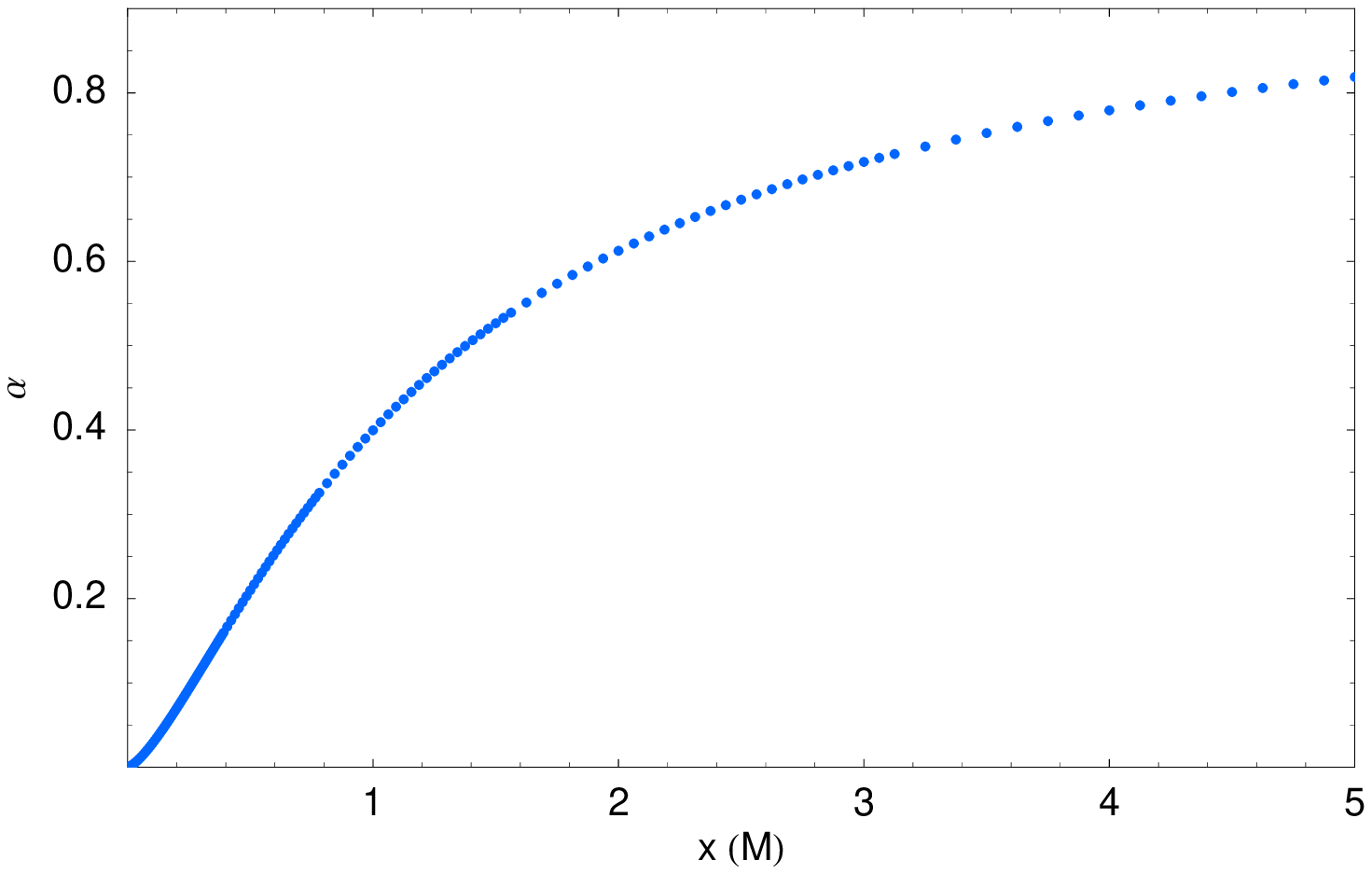}
\includegraphics[angle=0,clip,width=.45\textwidth]{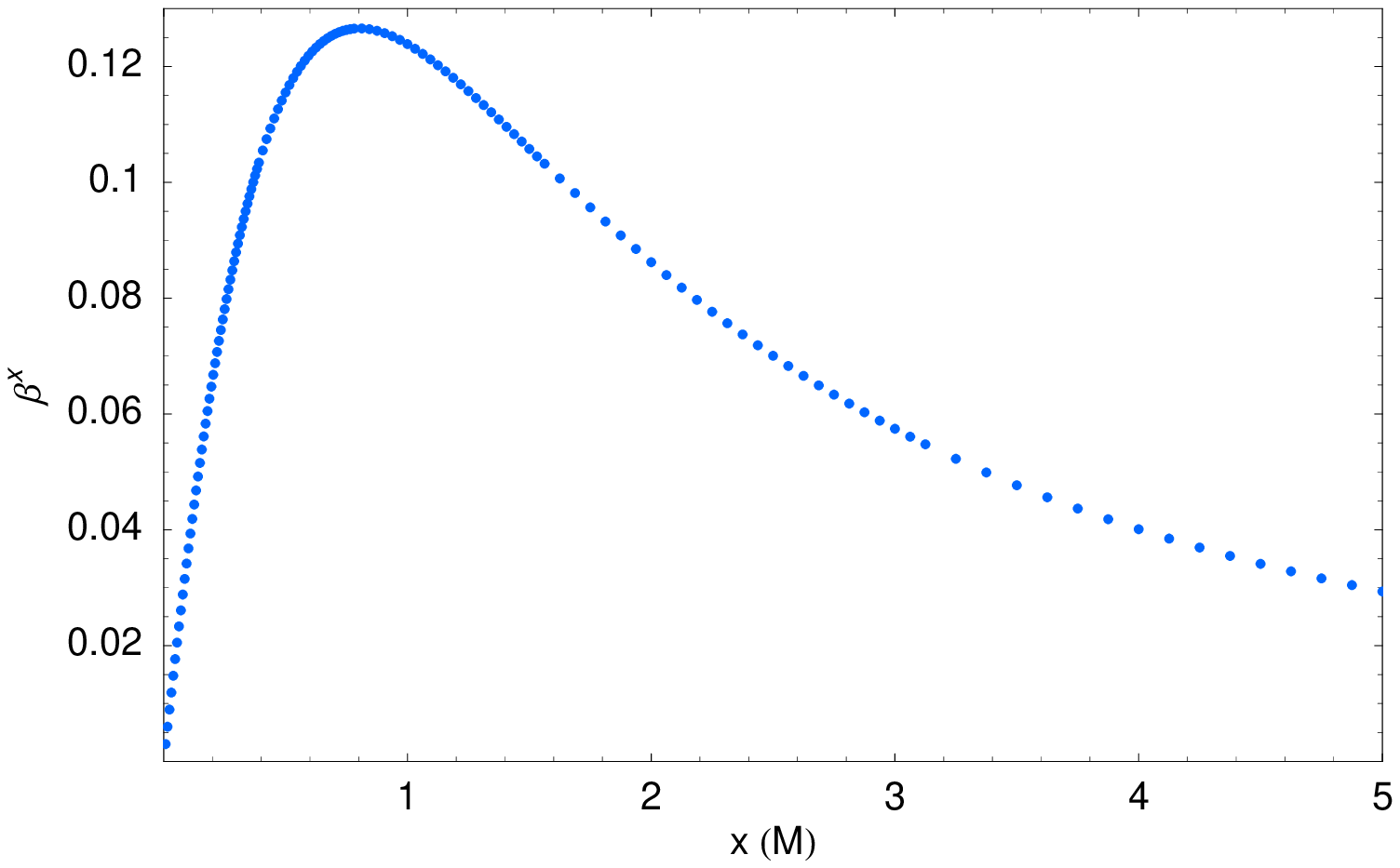}
\caption{The lapse function and $x$-component of the shift vector for
  time-independent puncture data for the Schwarzschild spacetime, based on the
  $t \rightarrow \infty$ limit of the Estabrook {\it et al} solution.}
\label{fig:cylinder}
\end{figure}

Having found the appropriate conformal factor $\psi$, we can reconstruct the 
background lapse and shift, and background extrinsic curvature
$\tilde{A}_{ij}$, and use these as initial data for a dynamical evolution. The
lapse and $x$-component of the shift vector constructed using
(\ref{eqn:EstaG})-- (\ref{eqn:EstaN}) and the numerical conformal factor
(\ref{eqn:NumPsi}) are shown in Figure \ref{fig:cylinder}. 

\begin{figure}
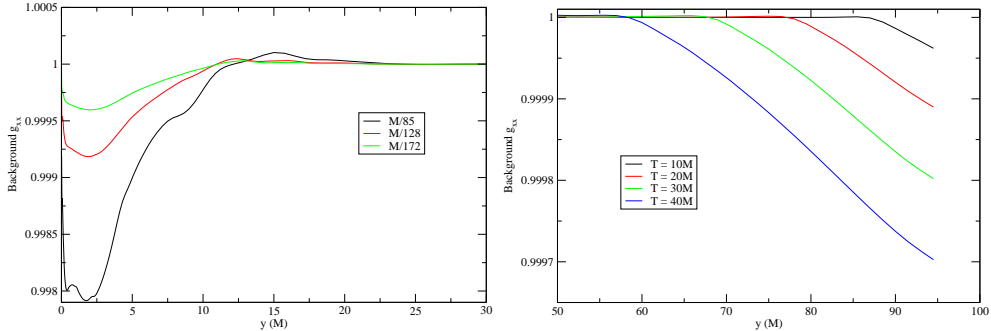

\centering
\includegraphics[angle=0,clip,width=.4\textwidth]{fig6a}
\includegraphics[angle=0,clip,width=.4\textwidth]{fig6b}
\caption{Errors in the evolution of the time-independent Schwarzschild initial
  data. The left panel shows the error in $\tilde{\gamma}_{xx}$ after $20M$ of
  evolution, with grid resolutions of $h = M/85,M/128,M/172$. The right panel shows
  the error that has propagated in from the outer boundary after $T = 10M,20M,30M,40M$.}
\label{fig:cylErrors}
\end{figure}

These data are indeed stationary: when we evolve them with the moving-puncture
method we see no obvious change in the grid functions with time. A calculation
of the error as a function of time (the difference between the initial data
and the evolved variables) shows that there {\it is} some nontrivial
evolution, but that this converges away with resolution. This is shown in
Figure \ref{fig:cylErrors}, where we see that numerical noise emerges from the
puncture. It is possible that this noise is largely due to the error in taking
finite-difference derivatives across the puncture of variables that are not
sufficiently smooth there. (For example, the lapse behaves like $\alpha \sim
|r|$ across the puncture.) However, the noise is small, it does not grow with
time, and it does not cause the simulation to crash. We also see noise
propagating in from the other boundary in the second panel in Figure
\ref{fig:cylErrors}, which does not converge away. This is due 
to the physically incorrect outer boundary conditions. 

These stationary data provide an excellent testbed for numerical evolutions:
knowing the correct analytic values of all evolution variables, we can
carefully monitor all errors in a simulation. It may also be possible to
generalize this construction to produce data of a similar type for black-hole
binaries. The physical quality of the data will depend on the choices of the
free data in our initial-data construction, but even if they do not have any
better physical properties than 
the Bowen-York data currently used in most moving-puncture simulations, they
would represent black holes in coordinates in which the initial gauge dynamics
would be minimized, in a similar fashion to the ``quasi-equilibrium'' conformal
thin-sandwich initial-data sets (see, for example, \cite{Cook:2004kt} and
\cite{Scheel:2006gg}).  

\section{Conclusion}

We have described the evolution of the geometry of puncture initial data in the
``moving puncture'' method in the context of asymptotically maximal slices of
the Schwarzschild spacetime. This description complements the picture in terms
of the 1+log slicing (\ref{eqn:logwithshift}) described in
\cite{Hannam:2006vv}. The main features of these
evolutions are that the numerical slices lose contact with the second
asymptotically flat end that is present in the initial data. The slices
instead end on a cylinder of finite Schwarzschild radial coordinate (but
infinite proper distance) inside the event horizon. As a result, the conformal
factor that initially diverged as $1/r$ at the puncture now has a milder
divergence of $1/\sqrt{r}$. The numerical evolution reaches a stationary state
in about $40M$, and the final stationary state can be described by the $t
\rightarrow \infty$ limit of the Estabook {\it et al} solution. Knowledge of
the analytic solution in turn allows us to construct initial data that is
unchanging in a numerical evolution, and provides a convenient experimental
environment to test the accuracy and stability of numerical codes. In
addition, these ``cylindrical data'' may provide insight into a method to
produce black-hole binary initial data in coordinates better suited to
numerical evolutions.

\section*{Acknowledgments}

We thank Denis Pollney for discussions. This work was supported in part by DFG grant
SFB/Transregio~7 ``Gravitational Wave Astronomy''. 
Computations where performed at HLRS (Stuttgart) and LRZ (Munich). We also
thank the DEISA Consortium (co-funded by the EU, FP6 project 
508830), for support within the DEISA Extreme Computing Initiative
(www.deisa.org).

\section*{References}

\bibliographystyle{iopart-num}
\bibliography{references_cvs,references_extra}

\providecommand{\newblock}{}
\begin{thebibliography}{10}
\expandafter\ifx\csname url\endcsname\relax
  \def\url#1{{\tt #1}}\fi
\expandafter\ifx\csname urlprefix\endcsname\relax\def\urlprefix{URL }\fi
\providecommand{\eprint}[2][]{\url{#2}}

\bibitem{Brill63}
Brill D~S and Lindquist R~W 1963 {\em Phys. Rev.\/} {\bf 131}(1) 471--476

\bibitem{Brandt97b}
Brandt S and Br{\"u}gmann B 1997 {\em Phys. Rev. Lett.\/} {\bf 78}(19)
  3606--3609

\bibitem{Dain02b}
Dain S 2002 {\em Lect. Notes Phys.\/} {\bf 604} 161--182

\bibitem{Bruegmann97}
Br{\"u}gmann B 1999 {\em Int. J. Mod. Phys. D\/} {\bf 8} 85

\bibitem{Alcubierre00b}
Alcubierre M, Benger W, Br{\"u}gmann B, Lanfermann G, Nerger L, Seidel E and
  Takahashi R 2001 {\em Phys. Rev. Lett.\/} {\bf 87} 271103

\bibitem{Reimann:2003zd}
Reimann B and Br{\"u}gmann B 2004 {\em Phys. Rev. D\/} {\bf 69} 044006

\bibitem{Reimann:2004pn1}
Reimann B and Brugmann B 2004 {\em Phys. Rev. D\/} {\bf 69} 124009

\bibitem{Campanelli:2005dd}
Campanelli M, Lousto C~O, Marronetti P and Zlochower Y 2006 {\em Phys. Rev.
  Letter\/} {\bf 96} 111101

\bibitem{Baker:2005vv}
Baker J~G, Centrella J, Choi D~I, Koppitz M and van Meter J 2006 {\em Phys.
  Rev. Lett.\/} {\bf 96} 111102

\bibitem{Pretorius:2005gq}
Pretorius F 2005 {\em Phys. Rev. Lett.\/} {\bf 95} 121101

\bibitem{Herrmann:2006ks}
Herrmann F, Shoemaker D and Laguna P 2006  (\textit{Preprint}
  \eprint{gr-qc/0601026})

\bibitem{Sperhake:2006cy}
Sperhake U 2006  (\textit{Preprint} \eprint{gr-qc/0606079})

\bibitem{Scheel:2006gg}
Scheel M~A {\em et~al.\/} 2006 {\em Phys. Rev. D\/} {\bf 74} 104006

\bibitem{Bruegmann:2006at}
Br{\"u}gmann B, Gonz\'alez J~A, Hannam M, Husa S, Sperhake U and Tichy W 2006
  (\textit{Preprint} \eprint{gr-qc/0610128})

\bibitem{Pollney:NFNR}
Pollney D July 2006 Presented at the {\it New Frontiers in Numerical
  Relativity} meeting, Golm

\bibitem{Hannam:2006vv}
Hannam M, Husa S, Pollney D, Br{\"u}gmann B and {\'O~Murchadha} N 2006
  (\textit{Preprint} \eprint{gr-qc/0606099})

\bibitem{Arnowitt62}
Arnowitt R, Deser S and Misner C~W 1962 in L~Witten, ed, {\em Gravitation: An
  introduction to current research\/} (New York: John Wiley) pp 227--265

\bibitem{York79}
York J~W 1979 in L~L Smarr, ed, {\em Sources of gravitational radiation\/}
  (Cambridge, UK: Cambridge University Press) pp 83--126 ISBN 0-521-22778-X

\bibitem{Bona97a}
Bona C, Mass{\'o} J, Seidel E and Stela J 1997 {\em Phys. Rev. D\/} {\bf 56}
  3405--3415

\bibitem{Alcubierre02a}
Alcubierre M, Br{\"u}gmann B, Diener P, Koppitz M, Pollney D, Seidel E and
  Takahashi R 2003 {\em Phys. Rev. D\/} {\bf 67} 084023

\bibitem{Shibata95}
Shibata M and Nakamura T 1995 {\em Phys. Rev. D\/} {\bf 52} 5428

\bibitem{Baumgarte99}
Baumgarte T~W and Shapiro S~L 1999 {\em Phys. Rev. D\/} {\bf 59} 024007

\bibitem{Hannam:2003tv}
Hannam M~D, Evans C~R, Cook G~B and Baumgarte T~W 2003 {\em Phys. Rev. D\/}
  {\bf 68} 064003

\bibitem{Estabrook73}
Estabrook F, Wahlquist H, Christensen S, DeWitt B, Smarr L and Tsiang E 1973
  {\em Phys. Rev. D\/} {\bf 7}(10) 2814--2817

\bibitem{Baumgarte:2007ht}
Baumgarte T~W and Naculich S~G 2007  (\textit{Preprint} \eprint{gr-qc/0701037})

\bibitem{Cook:2004kt}
Cook G~B and Pfeiffer H~P 2004 {\em Phys. Rev. D\/} {\bf 70} 104016

\end{thebibliography}

\end{document}